\input{aipcheck}

\documentclass[
   final            
  ]
  {aipproc}
\layoutstyle{6x9}

\newcommand{\be}{\begin{equation}}
\newcommand{\ee}{\end{equation}}

\newcommand{\eq}[1]{Eq. (\ref{#1})} 
 
\newcommand{\Fig}[1]{Fig. (\ref{#1})} 
 \newcommand{\eqa}{\begin{eqnarray}}
\newcommand{\eeq}{\end{eqnarray}}  
\newcommand{\eqsto}[2]{Eqs. (\ref{#1}) to (\ref{#2})}

\begin{document}

\title{ Passive Convection of Density Fluctuations in the Local Interstellar Medium}



\author{Dastgeer Shaikh and G. P. Zank}
{address={Institute of Geophysics and Planetary Physics,
University of California, Riverside, CA 92521.}}

\begin{abstract}
	We have developed a time-dependent three-dimensional model of
isotropic, adiabatic, and compressible magnetohydrodynamic plasma to
understand nonlinear cascades of density fluctuations in local
interstellar medium. Our simulations, describing evolution of initial
supersonic, super Alfv\'enic plasma modes, indicate that nonlinear
interactions lead to damping of plasma motion. During the process,
turbulent cascades are governed predominantly by the Alfv\'enic modes
and velocity field fluctuations evolve towards a state charachterized
by near incompressibility. Consequently, density field is advected
passively by the velocity field. Our findings thus demonstrate that
the observed density fluctuations in the interstellar medium are the
structures passively convected by the background velocity field.
\end{abstract}

\maketitle


\section{Introduction}

Density fluctuations in the local interstellar medium (ISM) are
observed to follow a Kolmogorov-like $k^{-5/3}$ spectrum \cite{kol}
that spans almost 12 decades in wavenumber space \cite{amstrong}.
This enigmatic observation has remained a source of inconspicuous
understanding of the origin, nature and dynamics of the ISM plasma
fluctuations. Interestingly, the Kolmogorov-like $k^{-5/3}$ spectrum
is a characteristic of fully developed incompressible hydrodynamic
turbulence. Why compressible and magnetized ISM plasma behaves like an
incompressible hydrodynamic fluid has been elusive and has certainly
obscured our understanding of small-scale ISM turbulence.  Small-scale
turbulent ISM fluctuations are not only potentially important in the
context of the global heliosphere, i.e. evolution of the termination
shock and its interaction with local small-scale upstream/downstream
turbulence, but are also instrumental to our understanding of many
astrophysical phenomena including energization and transport of cosmic
rays, gamma-ray bursts, ISM density spectra, etc.  To understand the
origin and dynamical evolution of turbulent density field in the local
ISM, we in this paper present our results of a self-consistent ISM
turbulence simulation model that we have devloped based on
time-dependent, three-dimensional magnetohydrodynamic equations.

\section{MHD Model}

Our model assumes that the ISM turbulent fluctuations in the plasma
are isotropic, homogeneous, thermally equilibrated and fully
developed. Second, no mean magnetic field and velocity flows are
present at the outset. There may however be local mean flows generated
by self-consistently excited nonlinear instabilities. We further
assume that the characteristic turbulent correlation length-scales are
typically much bigger the shock characteristic scale-lengths in the
ISM flows. Additionally, underlying turbulent correlation
length-scales are considered to be large enough to treat any localized
shocks as smooth discontinuities. In other words, the characteristic
shock length-scales are small compared to the ISM turbulent
fluctuation length-scales, and finally boundary conditions are
periodic, since we are not aware of any realistic boundary conditions
for the present problem. Periodicity is thus a natural and most
appropriate choice for modeling the local ISM.  Statistically
homogeneous, isotropic and isothermal MHD plasma can be cast in terms
of a single fluid density $\rho({\bf r}, t)$, magnetic ${\bf B}({\bf
r}, t)$ and velocity ${\bf U}({\bf r}, t)$ fields and pressure $p({\bf
r}, t)$ as
\be
\label{density}
\frac {\partial \rho}{\partial t} + \nabla \cdot (\rho {\bf U}) = 0, 
\ee
\be
\frac{\partial {\bf B}}{\partial t} = 
\nabla \times ({\bf U} \times {\bf B} ) + \eta \nabla^2 {\bf B},
\ee
\be
\label{mom}
 \rho \left(\frac{\partial}{\partial t} + {\bf U} \cdot \nabla \right){\bf U} = - \nabla p +
\frac{1}{4\pi}(\nabla \times {\bf B}) \times {\bf B}
\nonumber + \nu \nabla^2 {\bf U} 
+\hat{\eta} \nabla (\nabla\cdot {\bf U}).
\ee
\be
\label{velocity}
\nabla \cdot {\bf B}=0.  
\ee 
The equations are closed with an equation of state relating the
perturbed density to the pressure variables. Here ${\bf r}=x\hat{\bf
e}_x+y\hat{\bf e}_y+z\hat{\bf e}_z$ is a three dimensional vector,
$\eta$ and $\nu$ are, respectively, magnetic and kinetic viscosities.
Note carefully that MHD plasma momentrum equation, i.e. \eq{mom},
contains nonlinear dissipative terms on the right hand side ({\it
rhs}). This means that dissipative processes can potentially be
mediated by nonlinear interactions, in addition to the damping
associated with the small-scale turbulent motion. Thus nonlinear
turbulent cascades are not only responsible for the spectral transfer
of energy in the inertial range, but also likely to damp the plasma
motion in a complex manner. Nonetheless, the spatio-temporal scales in
the nonlinear damping can be {\it distinct} from that of the linear
dissipation.  We will quantify the damping associated with the
nonlinear interactions in the subsequent section.

\begin{figure}[t]
  \includegraphics[width=.5\textwidth]{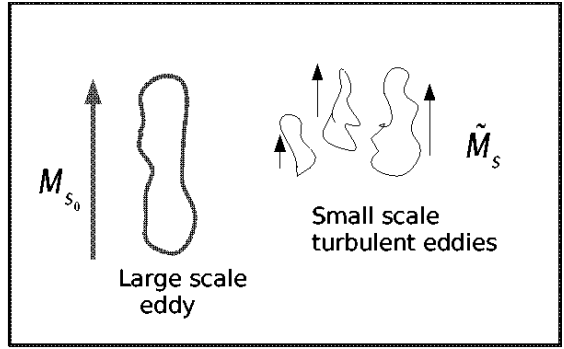} 
\caption{\label{fig0}
Schematic of the Mach number as determined from
the large-scale flows (left) and small-scale fluctuations (right). A
large-scale flow or constant mean background flow leads typically to a
constant Mach number, whereas local fluctuating eddies give rise to
turbulent Mach numbers which depend upon local properties of high
frequency and smaller-scale turbulent fluctuations.}
\end{figure}

The above equations can be normalized using a typical length scale
($\ell_0$), density ($\rho_0$), pressure ($p_0$), magnetic field
($B_0$) and the velocity ($U_0$). With respect to these normalizing
ambient quantities, one may define a constant sound speed $C_{s_\circ}
=
\sqrt{\gamma p_0/\rho_0}$, sonic Mach number $M_{s_\circ} =
U_0/C_{s_\circ}$, Alfv\'en speed $V_{A_\circ}= B_0/
\sqrt{4\pi\rho_0}$, and Alfv\'enic Mach number $M_{A_\circ} =
U_0/V_{A_\circ}$. The magnetic and mechanical Reynolds numbers are
$R_{m_\circ} \approx U_0 \ell_0/\eta$ and $R_{e_\circ} \approx U_0
\ell_0/\nu$, and the plasma beta $\beta_0 = 8\pi p_0/B_0^2$.  While
these quantities arise purely out of the normalizations and are
associated with a bulk (large-scale) plasma motion, there can exist
turbulent speeds, Mach and Reynolds numbers which depend locally on
the small-scale and relatively high frequency fluctuations. This is
illustrated schematically in \Fig{fig0}.  It is this component that
describes the high frequency contribution corresponding to the
acoustic time-scales in the modified pseudosound relationship proposed
in the Nearly Incompressible (NI) theory by
\cite{zank1990,zank1991,zank1993}.  Moreover, this high frequency component
is also related with the nonlinear damping of plasma motion as
described above.  We define the sound speed excited by the small-scale
turbulent motion as $\tilde{C}_s({\bf r}, t) =
\sqrt{\gamma} \rho^{(\gamma-1)/2}$, where 
$\gamma$ being the ratio of the specific heats, the sonic turbulent
Mach number $\tilde{M}_s({\bf r}, t) = \sqrt{\langle |{\bf
U}|^2\rangle}/\tilde{C}_s,$ and the fluctuating Alfv\'enic speed
$\tilde{V}_A = \tilde{B}/\sqrt{4\pi\tilde{\rho}}$, and the turbulent
Alfv\'enic Mach number $\tilde{M}_A({\bf r}, t) = \sqrt{\langle |{\bf
U}|^2\rangle}/\tilde{V}_A$.  The turbulent Reynolds numbers and plasma
beta $\tilde{\beta}$ can be defined correspondingly. We follow the
evolution of these volume-integrated local quantities in time to
understand the predominance of incompressible Alfv\'enic fluctuations
in the Solar wind and local interstellar medium which may be
responsible for turbulent cascades of energy. Furthermore, all the
small-scale fluctuating parameters are measured in terms of their
respective normalized quantities.

\section{Results and Discussion}

\begin{figure}[t]
  \includegraphics[width=.5\textwidth]{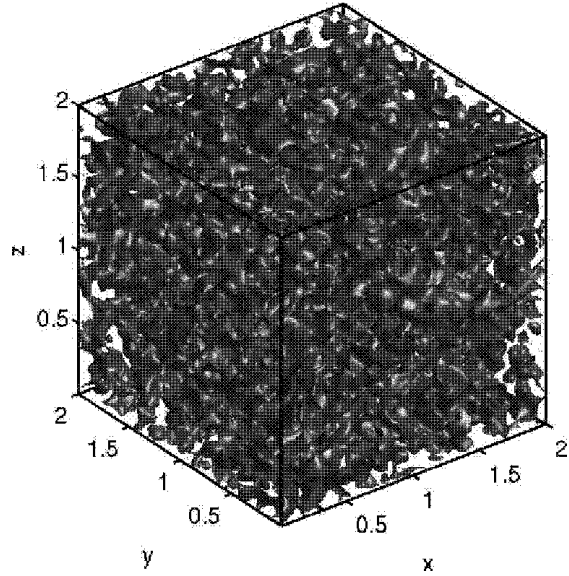} \caption{\label{fig1}
 Snap-shot of turbulent magnetic field in three dimensions. Shown are
 the iso-surfaces of $|{\bf B}|$ turbulent fluctuations at
 intermediate time step. The computation domain has volume of
 $(2\pi)^3$, spectral resolution is $128^3-256^3$. Initial turbulent
 Reynolds numbers are $\tilde{R}_e=\tilde{R}_m\approx 200$ (drop to
 about 20-25\%). Other parameters are $\gamma=5/3,
 \beta_\circ=10^{-3}, M_{A_\circ}=1.-2., M_{s_\circ}=5.-10.,
 \eta=\mu=10^{-4}-10^{-5}, dt=10^{-3}$.  Different resolutions and
 initial conditions do not qualitatively change the evolution
 exhibited in the figure.}
\end{figure}

We have developed a three dimensional compressible MHD code to
numerically integrate \eqsto{density}{velocity}.  The spatial
discretization in our code uses a discrete Fourier representation of
turbulent fluctuations based on a pseudospectral method, while the
temporal integration is performed by Runge Kutta 4 method. The
simulation parameters are described in \Fig{fig1}. All the
fluctuations are initialized isotropically (no mean fields are
assumed) with random phases and amplitudes in Fourier space.  This
algorithm ensures conservation of total energy and mean fluid density
per unit time in the absence of external random forcing. Our code is
massively parallelized using Message Passing Interface (MPI) libraries
to facilitate higher resolution. The initial isotropic turbulent
spectrum of solenoidal as well as irrotational velocity components was
chosen to be close to $k^{-2}$. It is to be noted, however, that our
results of turbulent cascades do not depend upon the choice of initial
spectrum. Therefore a flatter or steeper than $k^{-2}$ spectrum leads
qualitatively to similar results.  The ISM turbulence code is evolved
with time steps resolved self-consistently by the nonlinear
interaction time scales associated with the convective plasma motion,
i.e. $1/{\bf k} \cdot {\bf U}({\bf k})$.

ISM plasma relaxes in the absence of external sources and sinks.  As
plasma evloves, energy cascades amongst turbulent eddies of various
scale sizes in the ISM magnetoplasma. A snapshot of the $x$-component
of the fluctuating magnetic field in the ISM plasma is shown in
\Fig{fig1}.  During the evolution, MHD turbulent fluctuations are
dissipated nonlinearly gradually due to the finite Reynolds number in
which nonlinear interaction transfer spectral energy from the smaller
Fourier modes to the larger ones until turbulent excitation is
terminated by Kolmogorov dissipation modes ($k_d$). This is where
inertial range of the spectral cascades also terminates. The spectral
migration of energy among various modes in the inertial range leads to
a net decay of turbulent sonic Mach number $M_s$ as shown in
\Fig{fig2}.  The turbulent sonic Mach number continues to decay from a
supersonic ($M_s>1$) to a subsonic ($M_s<1$) regime. This indicates
that dissipative effects, triggered essentially by the nonlinear
interactions, predominantly cause the supersonic MHD plasma
fluctuations to damp strongly leaving primarily subsonic fluctuations
in the MHD fluid. Furthermore, intrigued by the prediction of the NI
theory, we monitor closely the evolution of the density fluctuations
and turbulent sonic Mach number in view of understanding a
relationship between the two, if there exists any. This is because the
small amplitude and small-scale density fluctuations in the NI theory
are shown to scale as $|\delta
\rho| \sim {\cal O} (M_s^2) \sim
\varepsilon$, where $\varepsilon= |\delta \rho|/\rho_0 \sim 10\%$ is treated as
a small parameter in the perturbative expansion of ISM or solar wind
velocity, magnetic field, pressure and density.  Interestingly enough,
we find a systematic development of the relationship between the
density fluctuations and turbulent sonic Mach number. This is shown in
\Fig{fig2} by  solid-curve.  Initially when the plasma is
supersonic, the turbulent sonic Mach number is large enough compared
to rms magnitude of the density fluctuations. The ratio of the density
fluctuations and turbulent sonic Mach number in the initial stage thus
shows a smaller value ($<10^{-2}$). As the ISM plasma evolves, the
nonlinear interactions lead to subsonic motions of turbulent modes
which are characterized by a smaller magnitude of $M_s$
i.e. $M_s<1$. Since there are no damping processes involved in the
evolution of the density fluctuations, they do not dissipate. By
contrast, the density fluctuations begin to order quadratically with
the turbulent Mach number. This consequently leads to the relationship
$|\delta \rho| / (M_s^2) \sim {\cal O} (1)$ in our 3D simulations, as
shown in
\Fig{fig2} (the solid curve). Note that turbulent motion 
of the velocity field fluctuations associated with the smaller Mach
number in the ISM is regarded predominantly as incompressible. This,
in agreement with most observations
\cite{amstrong,zank1990,zank1991,zank1993}, points towards the fact
that the ISM velocity field fluctuations are composed predominantly of
(nearly) incompressible component.  As a direct consequence of the ISM
magnetoplasma being nearly incompressible, the density fluctations
exhibit a weak compressibility in the gas and are convected
predominantly passively in the background incompressible fluid flow
field. This hypothesis can also be verified straightforwardly by
investigating the density spectra which should be slaved to the
incompressible velocity spectra \cite{dastgeer}.

\begin{figure}[t]
  \includegraphics[width=.5\textwidth]{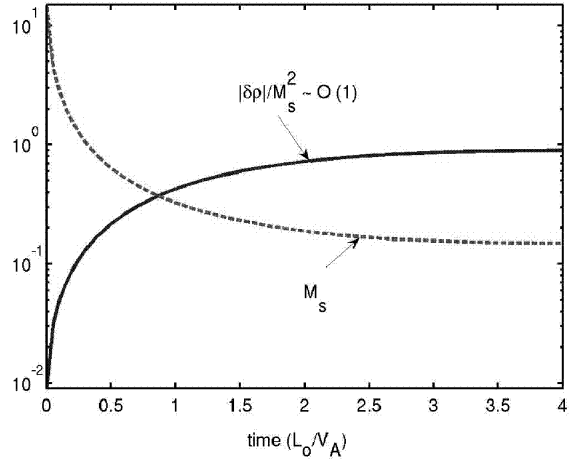} \caption{\label{fig2r}
 Snap-shot of turbulent magnetic field in three dimensions. Shown are
 the iso-surfaces of $|{\bf B}|$ turbulent fluctuations at
 intermediate time step.}
\end{figure}

The transition of compressible magnetoplasma from a(n) (initial)
supersonic to a subsonic or nearly incompressible regime is
gradual. This means the magnetofluid contains supersonic, and super
Alfv\'enic modes initially in which highly compressible density
fluctuations do not follow the velocity spectrum. It is the eventual
decay of the turbulent Mach number to a subsonic regime that is
responsible for the density fluctuations following the velocity
fluctuations. In the subsonic regime, the compressibility weakens
substantially so that the density fluctuations are advected only
passively.  A passively convected fluid follows a similar inertial
range spectrum as that of its background flow field
\cite{macomb}. Likewise, subsonic density fluctuations are also expected
to exhibit a spectrum similar to the background velocity fluctuations
in the inertial range, as demonstrated in our work \cite{dastgeer}.

An alternate understanding of the passive scalar evolution of the
density fluctuations associated essentially with incompressiblity can
also be elucidated directly from the continuity equation as follows.
\be
\label{cont}
(\partial_t + {\bf U} \cdot \nabla) \ln \rho = - (\nabla
\cdot {\bf U}). 
\ee
The rhs in \eq{cont} represents compressiblity of the velocity field
fluctuations. It is clear that the density field is advected passively
through the velocity field of the fluid when it is incompressible,
i.e. when solenoidal component is vanishingly small. This leads to the
expression of the solenoidal velocity field as $\nabla \cdot {\bf U}
\simeq 0$, and fluid continuity equation becomes $(\partial_t + {\bf
U} \cdot \nabla) \ln \rho \simeq 0$.

\section{Conclusion}
In conclusion, we have developed a self-consistent nonlinear 3D MHD
fluid model to describe turbulent cascade processes that lead to a
passively convected density fluctuation spectrum in the local
interstellar medium and/or solar wind plasma. In an undriven
situation, we find that an initial supersonic, compressible MHD plasma
relaxes towards a state that comprises predominantly of subsonic and
(nearly) incompressible plasma motion by virtue of suppressing the
solenoidal velocity field fluctuations. The suppression is mediated
explicitly by means of nonlinear interactions in which transverse
Alfv\'enic fluctuations do not couple with co-existing longitudinal
fast/slow modes. In view of the weak interaction between the two
competing MHD modes, the nonlinear cascades are dominated by the
Alfv\'enic fluctuations
\cite{dastgeer1}. The latter is entirely responsible for generating
the density fluctuations as passively convected structures or {\it
pseudo modes} (i.e. non interacting modes). This has also been
demonstrated by us in a recent 3D compressible MHD simulations of ISM
plasma \cite{dastgeer1}.  One of the important implications of the
passive advection of the density fluctuations is that their
characteristic spectrum can be determined from the background (nearly)
incompressible velocity fluctuations. This strong correlation, is
perhaps, responsible for the observed turbulent density spectrum in
the local interstellar medium \cite{amstrong}. We finally point out
that our results will not differ in the presence of external driving
forces such as those arising from large-scale instability, supernovea
and other possible sources as long as solenoidal component of the
velocity field fluctuations is not affected dramatically to supersede
the Alfv\'enic cascades.



\bibliographystyle{aipproc}   

\begin{thebibliography}{}

\bibitem{kol}  
A. N. Kolmogorov,   Dokl. Acad. Sci. URSS, {\bf 30}, 301, 1941.

\bibitem{amstrong} 
J. W. Armstrong, J. M. Cordes, and B. J. Rickett,
{\bf 291}, 561 1981.


\bibitem{zank1990} 
G. P. Zank, and W. H. Matthaeus,
Phys. Rev. Lett., {\bf 64}, 1243, 1990.

\bibitem{zank1991}
G. P. Zank, and  W. H. Matthaeus,
Phys. Fluids A,  3, 69,  1991.

\bibitem{zank1993} 
G. P. Zank, and W. H. Matthaeus,
Phys. Fluids, A {\bf 5}, 257, 1993.

\bibitem{dastgeer} 
D. Shaikh and G. P. Zank, Manuscript in preparation, 2007

\bibitem{dastgeer1} 
D. Shaikh and G. P. Zank, Astrophys. J., 625, 2006.

\bibitem{macomb}
W. D. McComb, Physics of Fluid Turbulence, Oxford Science
Publications, 1990.



\end{thebibliography}


\end{document}